\documentclass{article}

\usepackage{PRIMEarxiv}
\usepackage{amsmath}
\usepackage[utf8]{inputenc} 
\usepackage[T1]{fontenc}    
\usepackage{hyperref}       
\usepackage{url}            
\usepackage{booktabs}       
\usepackage{amsfonts}       
\usepackage{nicefrac}       
\usepackage{microtype}      
\usepackage{lipsum}
\usepackage{fancyhdr}       
\usepackage{graphicx}       
\graphicspath{{figures/}}     
\usepackage{natbib}
\usepackage{abstract}
 
\usepackage{setspace}

\newcommand{\citeg}[1]{\citep[e.g.,][]{#1}}

\title{How do applied researchers use the Causal Forest? A methodological review
}

\author{
  Patrick Rehill\\
  POLIS: The Centre for Social Policy and Research \\
  Australian National University \\
  Canberra, Australia\\
  patrick.rehill@anu.edu.au \\
  146 Ellery Crescent, Canberra ACT 2600 \\
}

\begin{document}
\maketitle
\doublespacing

\begin{abstract}
This methodological review examines the use of the causal forest method by applied researchers across 133 peer-reviewed papers. It shows that the emerging best practice relies heavily on the approach and tools created by the original authors of the causal forest such as their grf package and the approaches given by them in examples. Generally researchers use the causal forest on a relatively low-dimensional dataset relying on observed controls or in some cases experiments to identify effects. There are several common ways to then communicate results --- by mapping out the univariate distribution of individual-level treatment effect estimates, displaying variable importance results for the forest and graphing the distribution of treatment effects across covariates that are important either for theoretical reasons or because they have high variable importance. Some deviations from this common practice are interesting and deserve further development and use. Others are unnecessary or even harmful. The paper concludes by reflecting on the emerging best practice for causal forest use and paths for future research.
\end{abstract}


\section{Introduction}
Causal machine learning methods offer a flexible way to estimate
heterogeneous treatment effects nonparametrically by estimating conditional average treatment effects (CATEs). This is particularly
useful for high-dimensional datasets, those where drivers of
heterogeneity are not known in advance and those with nonlinear effects.
One of the most popular methods is the causal forest \citep{wager_estimation_2018,athey_generalized_2018}. It gives strong
`plug-and-play' performance for applied researchers who may not have
much experience designing and tuning machine-learning models \citep{wager_estimation_2018}. The most popular implementation is in the grf
package \citep{athey_generalized_2018} which includes some useful quality of life features like
auto-ML via a tuning forest, automatic fitting of nuisance models when
estimates are not provided, efficient estimation of standard errors via a bootstrapping of little bags, a simple implementation of double machine learning for average treatment effect (ATE) inference, and tools to help understand results.

In the years since the method was proposed many applied
papers have made use of it, however, the precise way they use it
varies. The aim of this paper is to try and understand the state of
applied work using the causal forest. This is a kind of
methodological review but focused not on one field of research, but on one method.
It examines peer reviewed papers indexed by Scopus and Web of
Science that have used these methods in applied work and tries to make sense of just how
they were used. This method is immensely promising for applied
researchers, but it is only just beginning to be applied. That is
why it is important to pay attention to the way the causal forest is being  applied,
highlight best practice, point out mistakes that have been made and
suggest directions the development of the causal forest may take in the future. 

The review covers 133 peer-reviewed publications across many different fields. It focuses on three broad concerns --- the identification of effects, the estimation of effects and the presentation of results. In the causal inference methods literature, how results of analysis with a method are presented generally gets much less attention than identification and estimation. However, the causal forest's main output (an often high-dimensional conditional average treatment effect distribution) is so challenging to interpret on its own that presentation of results deserves a special focus here. We need some way to digest such a distribution into something that provides useful insights, and there is not a canonical way to do so \citep{rehill_transparency_2023}.

This review finds that the causal forest is a method that is growing in use and which is being used in interesting and valuable ways across many different fields. However, there are also problems. The way the causal forest is used in applied work has been largely stagnant since the early Athey, Wager and co-authors papers. In the way effects are identified, estimated, and insights are extracted out of the model, researchers largely follow the practice of \citet{athey_generalized_2018}. What deviations there have been from this practice are as likely to be bad ideas as good ones. That is not to say that it is best to stick
to the current common practice. There are certainly
innovations in the methods literature which will hopefully filter into applied work in time.
There are also some successful innovations amongst these papers that
show what is possible when the methods are used in a way that is both
rigorous and creative.

\section{A brief introduction to the causal forest}
As this study will be talking a lot about the specificities of different causal forest designs, it is worth explaining what the causal forest is and how it works for any reader who may be unfamiliar. The causal forest is an approach to estimating CATEs based on the random forest --- a popular predictive machine learning method \citep{breiman_random_2001}. The method allows for asymptotically normal estimates to be made of CATEs using a matrix of predictors that can range from a few variables, to very high-dimensional datasets with more features than observations. In simulations, the method performs well estimating effects both in the presence of confounding, and complex treatment effect heterogeneity \citep{dorie_automated_2019, jawadekar_practical_2023}.

The standard causal forest has developed through several key methodological papers by Athey, Wager and co-authors \citep{athey_recursive_2016,wager_estimation_2018, nie_quasi-oracle_2020,athey_generalized_2018}. Its estimates are asymptotically unbiased and normal. There are many other variations on the causal forest that have been proposed for example by \citet{lechner_modified_2019} or \citet{dandl_what_2022}. The review will show that they are used little in practice so for simplicity I do not discuss them in depth here.

The causal forest has several key steps --- it removes confounding with nuisance models, it fits a model to cluster observations with treatment effect heterogeneity to form an adaptive kernel and then it estimates out treatment effects within each kernel bandwidth. Three models are used, a nuisance model that predicts treatment, a nuisance model that predicts outcome and finally a model of treatment effect heterogeneity. These nuisance models are generally random forests themselves (typically regression forests in grf) that can be used to partial out selection effects (similar to how they are used in double machine learning \citep{chernozhukov_doubledebiased_2018}). This partialing out is called local centering --- essentially taking the residuals of the models as the treatment and outcome variables used in the final model. The final heterogeneity model is a random forest made up of causal trees \citep{athey_recursive_2016}. These differ from standard regression trees in two ways. Firstly they use a special splitting criterion designed to find treatment effect heterogeneity (estimated expected mean squared error). Secondly they are 'honest', this means each tree uses typically half of its sample to split the tree and the other half to make estimates for each terminal node. This prevents over-fitting, giving asymptotically normal predictions out of each tree. The overall ensemble these fit into is trained as an R-Learner meta-learner \citep{nie_quasi-oracle_2020}. This means it minimises the R-Loss function to estimate a heterogeneity model $\tilde{\tau}$ as $$
\tilde{\tau}(\cdot)=\operatorname{argmin}_{\tau}\left(\frac{1}{n} \sum_{i=1}^{n}\left[\left\{Y_{i}-\hat{m}^{(-i)}\left(X_{i}\right)\right\}-\left\{W_{i}-\hat{e}^{(-i)}\left(X_{i}\right)\right\} \tau\left(X_{i}\right)\right]^{2}+\Lambda_{n}\{\tau(\cdot)\}\right).
$$Here we see the local centering ($Y_{i}-\hat{m}^{(-i)}(X_i)$, $W_{i}-\hat{e}^{(-i)}(X_{i})$) which gives us the residuals (sometimes called pseudo-outcomes) that the final model is fit on. These are fit out-of-bag (hence the $(-i)$ superscript. There is also a regularisation that is in this case, implicit in the structure of the causal forest (e.g. from tuning minimum node size $\Lambda_{n}\{\tau(\cdot)\}$). This loss function allows for the tuning of the forest automatically via a smaller tuning forest that is fit as part of the causal forest function.

The last innovation of the causal forest is that rather than predicting directly with an average of trees like a standard random forest, the method simply uses the ensemble as an adaptive kernel, finding weights to calculate treatment effects for neighbourhoods in a high-dimensional covariate space. We find the weights for a covariate set $x$ by looking through each data-point $i$ and finding the proportion of trees that were not fit on $X_i$ (to prevent overfitting) for which $X_i$ and $x$ fall into the same leaf. This gives us the kernel function $\alpha(x)$. More formally for each tree $b$ in the ensemble $B$ in the tree's sample $\mathcal{S}_b$,

$$\alpha_i(x) = \frac{1}{B} \sum_{b=1}^B \frac{\mathbf{1}\left( \{ X_i \in L_b(x), \, i \in \mathcal{S}_b \} \right)}{\left| \{ i : X_i \in L_b(x), \, i \in \mathcal{S}_b \} \right|}.
$$

In order to estimate treatment effects we use a weighted version of Robinson's transformation \citep{robinson_root-n-consistent_1988} as proposed by \citet{athey_generalized_2018}. $$\hat{\tau}(X_i) = \frac{\sum_{i=1}^n \alpha_i(x) \left( Y_i - \hat{m}^{(-i)}(X_i) \right) \left( Z_i - \hat{e}^{(-i)}(X_i) \right)}{\sum_{i=1}^n \alpha_i(x) \left( Z_i - \hat{e}^{(-i)}(X_i) \right)^2}
$$

Standard errors can be calculated via resampling, generally (i.e. in the grf approach) via a bootstrap of little bags. This leverages the random sampling that already takes place in fitting a random forest which allows for error estimation that is much less computationally intensive than a standard non-parametric bootstrap. One problem with just a standard resampling is that in small ensembles (where the number of trees is small) the variance estimate can turn negative. In order to avoid this a Bayesian analysis with a non-informative prior is used. This prevents negative estimates in small ensembles while having little effect on estimates for large ensembles \citep{athey_generalized_2018}.

This approach overall yields an asymptotically unbiased and normal estimator that can be fit easily on a researcher's personal computer and that does not require extensive manual hyperparameter tuning.

\section{Study design} 
The inclusion criteria for this
study were simple. The paper should use the causal forest
method and it should be primarily an applied paper, not a methods one.
The reason for this is simply that methods papers, even when they use
non-synthetic data are fundamentally a different kind of paper with
different aims, different methodologies, and different findings they want
to communicate. For example many were primarily interested in comparing the
causal forest against other methods and so emphasised the
goodness-of-fit of the model rather than the actual insights that
could be extracted.

The first step was searching Scopus and Web of Science using a keyword
search for "causal forest". This
yielded 197 unique articles. These were then screened based on their
abstracts which yielded 107 articles that were then read in full. 85 of
these were found to meet the inclusion criteria and were analysed in
this study.

A broader search of any paper citing \citet{wager_estimation_2018} on Scopus or Web of Science was then conducted. As any papers here that were not duplicates from the key word search would by definition not refer to the causal forest in their abstract, reading abstracts was not a feasible screening strategy. For this reason I attempted to retrieve all of these articles but rather than reading them in full, I first parsed them with a script. This searched the text for mentions of "causal forest" and provide the line of text around the mention for context so I could quickly assess why the Wager and Athey paper was cited. The vast majority of the papers cut here (keeping in mind these were papers that were not found in the key word search) mentioned but did not use the method, or cited the paper as part of a larger point about machine learning methods in causal inference. The next most common reason for exclusion was that papers were primarily methodological. 62 of the papers from Scopus possibly met the inclusion criteria and were read in full. This approach was not used for the Web of Science papers as there were only 23 left at this stage (because this search was conducted after the Scopus search meaning any duplicates were assessed as part of the Scopus sample or key word search). For this reason, the Web of Science papers were read in full. Only 40 Scopus papers and eight Web of Science papers that had not already been picked up in the key word search met the inclusion criteria.

\begin{figure}[!h]
\centering
\includegraphics[]{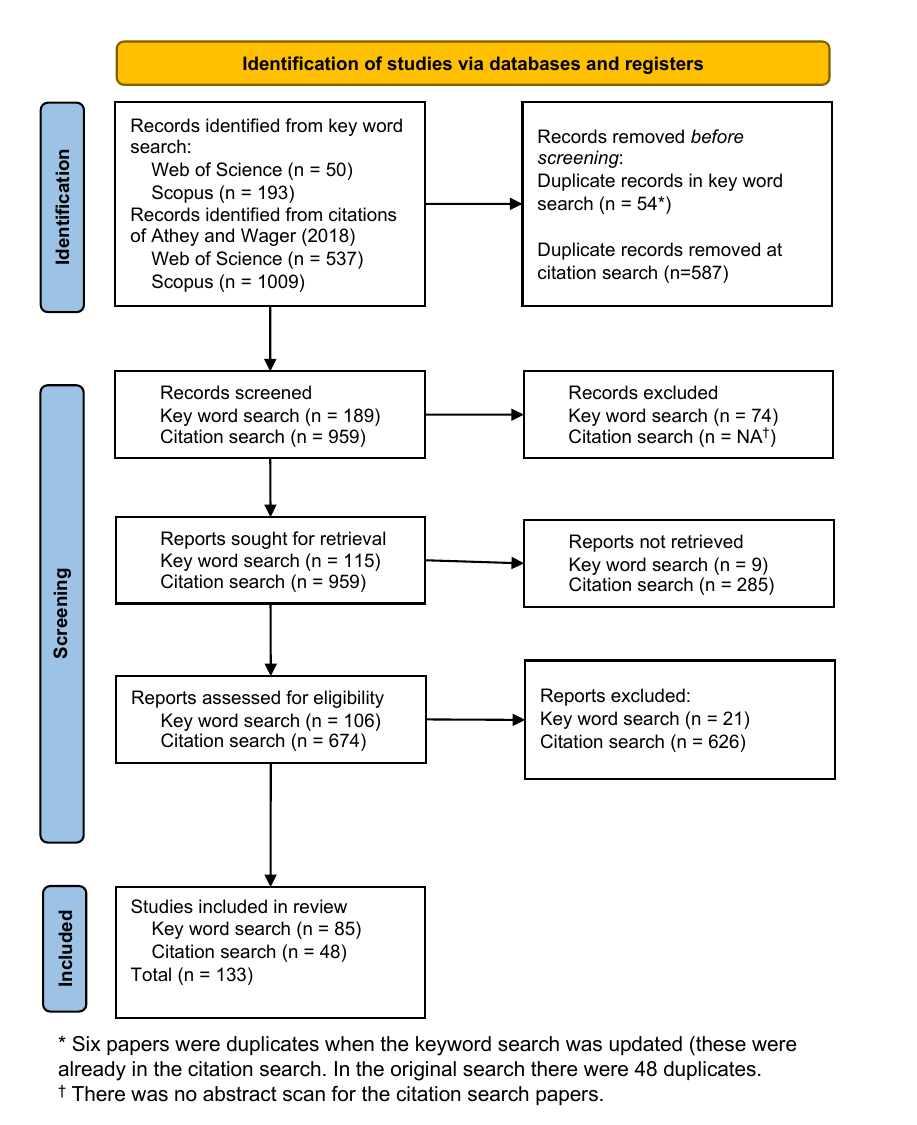}
\caption{Process for selecting studies into this review. Diagram from template by \citet{page_prisma_2021}.}
\label{fig:figure1}
\end{figure}

The citation search was conducted after the keyword search yielded 75 papers. Directly after the citation search, the keyword search results were updated and a further 10 papers had been published in that time that were not already included in the citation search. Figure \ref{fig:figure1} provides the full breakdown of this selection process. A full list of the papers selected and notes on their use of the causal forest is available in the web appendix.

\begin{figure}[h]
\centering
\includegraphics[width=.6\linewidth]{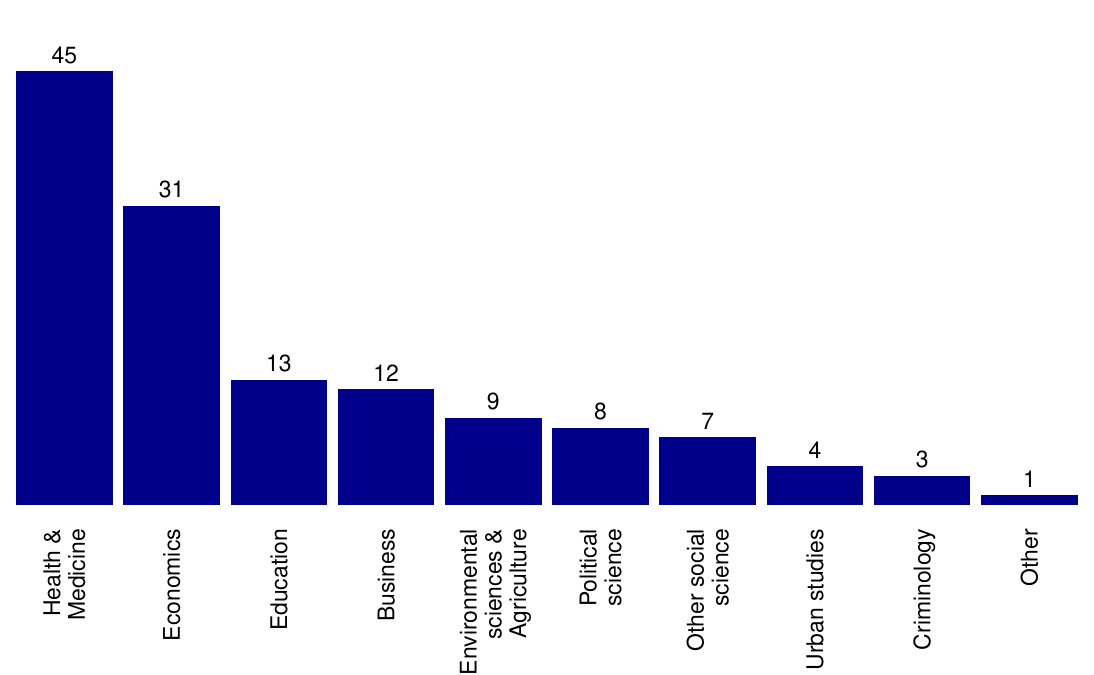}
\caption{Publications in study by their discipline}
\label{fig:figure2}
\end{figure}

\begin{figure}[h]
\centering
\includegraphics[width=.6\linewidth]{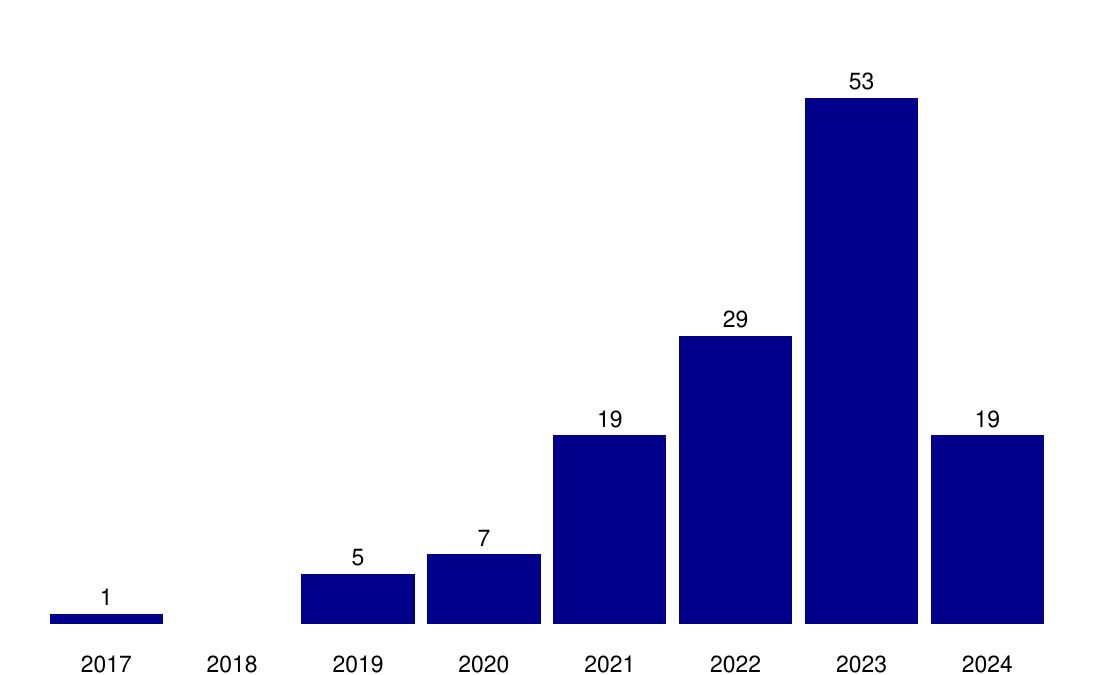}
\caption{Publications in study by year (note the search was conducted in April 2024 so only accounts for publications in the year to date at that point.)}
\label{fig:figure3}
\end{figure}

To provide context on the literature, I will briefly present some basic exploratory
analysis on the dataset. Figure \ref{fig:figure2} plots the number of publications by field while Figure \ref{fig:figure3} plots the distribution of papers over time.
These categorisations were made manually when reading the papers. It is clear that a
few disciplines represent a large part of the causal forest literature,
particularly economics and health research which together
account for most of the papers in the sample. Use of the method has grown over time. It particularly took off after the grf package and generalised random forest paper in 2019.

\section{Results - Identification and estimation approaches}

\subsection{Choice of software}

Across the sample a clear majority of papers that detailed the software they used used the grf package by \citet{athey_generalized_2018} to fit their causal forest. In fact only seven papers that name the software they used did not use grf. The ubiquity of grf affects the approach to identification, estimation and presentation of
results because some processes are easier than others with the tooling it offers and some have already been modelled in tutorials. Table \ref{tab:table1} below shows the (non-exclusive) count of the packages used
across different papers, i.e. if an approach combines two packages, it is counted once for each package. It is notable that despite the popularity of the EconML Python package in industry \citep{syrgkanis_causal_2021}, it was only explicitly used by one paper in the sample. As a proxy for use in industry we can look to GitHub stars. EconML has 3817 stars compared to grf's 968 stars. For context, the most starred CATE estimation library seems to be the CausalML library created at Uber.

While comparing use of the causal forest in a robust way with other techniques is out of the scope of this paper, it is worth noting that in general, the causal forest seems to be a solution better suited for the work researchers do than industry users. Companies with large amounts of data, that might come in sequentially, existing predictive machine learning pipe-lines, the ability to A/B test predictions (rather than relying on error distributions), and a staff of machine learning experts are much more likely to want to use something like a meta-learner employing deep learning or a boosting model. On the other hand, researchers with less data, who require good standard errors, who are unlikely to have in-depth experience in fitting and tuning machine learning models are more likely to benefit from a causal forest.

There is also a language divide as only six
paper used any Python packages at all \citep{kristjanpoller_causal_2021, seitz_individualized_2023} with the MCF package being used by three of these \citep{zhu_effect_2023,cockx_priority_2023,hodler_institutions_2023}.
Either because of the familiarity of academics with R or the quality of
the grf tooling, it appears R is simply dominant in causal forest
modelling. One paper used grf via the MLRtime Stata package.

\begin{table}[h!]
    \centering
    \begin{tabular}{ll}
        Name & Count\\
        grf & 69\\
        mcf \citep{lechner_modified_2019} & 3\\
        causalTree & 3\\
 EconML&1\\
        tools4uplift & 1\\
        causalML & 1\\
 Original causal forest implementation (from code associated with \citet{athey_recursive_2016})& 1\\
        MLRtime & 1 \\
        No paged stated & 54\\
    \end{tabular}
    \caption{Count of packages used in the papers studied (some use multiple packges).}
    \label{tab:table1}
\end{table}

While it is hard to say, my suspicion is that most of the papers that did not specify an implementation used the grf
package as they often cited the \citet{athey_generalized_2018} without naming an
implementation and they were generally using R where grf is the only popular implementation of a standard causal forest. In some cases with particularly novel designs or papers which predate the grf release \citeg{baum_targeting_2017},
it may be that the researchers simply implemented their own version or
took an earlier implementation (like that in Athey's causalTree
package available at https://github.com/susanathey/causalTree).

\subsection{Identification strategies} \label{sec:id-strat}
Most papers in the study used either experimental (54) or control on observables (72) designs. However there were a number of papers that
used interesting variations on these identification strategies.

\subsubsection{Controlling on observables}
It is interesting to see control-on-observables designs used so commonly, particularly in fields (like economics) where they are not generally favoured. Is there something about using powerful machine learning nuisance functions that allows us to identify causal effects without some kind of randomisation in a way we could not with simpler methods? Clearly many authors think there is and they are eager to use the nuisance models to identify effects. Whether they are right or wrong in each individual case is unanswerable for specific applied questions due to the Fundamental Problem of Causal Inference \citep{holland_statistics_1986}. In simulations, the causal forest seems robust to confounding when confounders are observed \citep{jawadekar_practical_2023, dorie_automated_2019}, but we of course cannot guarantee confounders are observed in real-world problems.  More emphasis on robustness checks or perhaps bounding error from unobserved confounders might be useful, but it is hard to know how to adapt methods from the ATE estimation literature to the problem of potentially thousands or millions of individual-level estimates. More work needs to be done to understand the credibility of such designs. Certainly wider use of robustness checks and in particular placebo tests would be useful.

Some of the papers pursue unusual routes to identifying observational effects; these often seem redundant. For example \citet{inoue_heterogeneity_2023} calculate propensity scores with logistic regression before matching
observations and then running a causal forest analysis which includes a propensity score weighting step with different scores calculated by a presumably more powerful generalised random forest model. At best this is redundant and at worst it may lead to problems estimating standard errors using resampling as the same limitations that apply to bootstrapping likely also apply here \citep{abadie_large_2006}.

When it comes to validating the assumptions under ignorability papers test both overlap and unconfoundedness in different ways. 23 papers use some kind of an overlap test, generally graphing propensity scores and seeing if there is sufficient overlap. In some cases researchers trim scores to ensure better overlap. Only a minority of control-on-observables papers explicitly do these tests (others could have done the tests but not reported them). Control-on-observables involves strong assumptions, and checking overlap is crucial for credible identification. Checking this assumption for the whole of the dataset should be a basic part of any control-on-observables design.

18 papers use some kind of falsification test for unconfoundedness. In all cases this is some kind of placebo treatment or dummy outcome test where a model is fit on a random synthetic treatment, a random synthetic outcome or a logically unrelated outcome (such as a pre-treatment period outcome). This is a standard approach with other methods as well and it seems to be effective here. However, it is interesting that a wider range of refutation tests like those discussed by \citet{sharma_dowhy_2021} are not used. It is unclear if there would be value to using some more of these when using the causal forest and if any of these tests would better suit CATE estimation.

One problem with most of these tests is that they are mostly conducted at the level of the average effect because it is much easier to test something once at the aggregate level than once for each CATE estimate. It is hard to know how much validation of ignorability at the highest level reflects ignorability in CATE estimates in practice. Certainly it is a daunting prospect to validate ignorability for each individual-level CATE estimation, but this is likely not neccessary. As the estimates out of a causal forest are only as useful as the ability to understand them \citep{rehill_transparency_2023}, it is only necessary to validate the assumptions at the level at which the estimates are being interpreted. This depends a lot on the specific methods used to present CATE results (discussed in more detail in Section \ref{results---presentation-of-results}). For example, if a researcher is dividing estimates into quantiles, the assumptions should hold in each quantile, if fitting a new, interpretable tree, the assumption should hold in each 'leaf' of the tree.

\subsubsection{Experiments}
It is notable that the causal forest from the grf package is being used so widely with experimental data. While the use of the causal forest on experimental data has been discussed in relation to other methods (the original causal forest \citep{wager_estimation_2018}, the model-based forest \citep{dandl_what_2022}), it has not been a focus of methodological discussion when using a forest with local centering. For most of the papers in this review, it is not clear whether oracle propensities were used, or whether propensities were estimated as if the data were observational. Just seven of the experimental papers detail which approach they used in the text of the paper and all of these use estimated scores rather than oracle propensity scores. It is not clear there is any benefit to using estimated scores. With inverse probability weighting per \citet{hirano_efficient_2003}, estimation of propensity scores with a nonparametric model compared to just using oracle propensities can provide a more efficient estimator of an average treatment effect at the cost of a small amount of bias. Therefore this approach is often favoured \citep{wager_causal_2024}. However with the orthogonal moment condition the causal forest uses it is likely that oracle scores are the superior approach \citep{grf_issue_636}. This is definitely the case when AIPW estimates are made as well \citep{wager_causal_2024}.

Three natural experiments or randomised controlled trials had data that was imperfectly randomised which provided some level of exogeneity to treatment but did not provide full identification \citep{brock_discriminatory_2023, allen_assessing_2022, habel_effective_2023, hodula_cooling_2023}. Adjusting with controls was still necessary to get identification but the partial randomisation provided a level of confidence in identification. One paper used a data fusion
design to combine experimental and observational data to improve
identification \citep{kluger_combining_2022}.

\subsubsection{Quasi-experimental}
Given how prevalent quasi-experimental approaches are in economics, and how many of these papers are from that discipline, it is curious how few quasi-experimental designs were used. Of course there is no in-built support for many common quasi-experimental methods in grf, but there is a built in instrumental variable method --- the instrumental forest \citep{athey_generalized_2018, tibshirani_package_2021}. Interestingly only one paper \citep{brooks_assessing_2022} used this method.

Six papers used some kind of
difference-in-differences design \citep{xue_local_2023, cui_tax-induced_2022, wang_befriended_2023, wang_effect_2022, miao_effects_2023, guo_effect_2021}. Notably, many others used a standard difference-in-differences with a linear fixed effects model design for their analysis of ATE but used control-on-observables identification with a causal forest to explore heterogeneity.

Generally where papers used difference-in-differences for treatment effect heterogeneity they used the first difference between the last
pre-treatment period and the first treatment period as an outcome and
then modelled the relationship with a standard causal forest. This is an approach that fits with the \citet{abadie_semiparametric_2005} work on semi-parametric difference-in-differences although strictly speaking simply pre-processing data and using a standard causal forest targets the CATE where a conditional average treatment effect on the treated (CATT) may be more appropriate for the same reasons most difference-in-differences designs target the average treatment effect on the treated (ATT). \citet{miao_effects_2023} formalises this design as the First Differences Causal Forest. This
approach helps to improve the credibility of the causal forest, at least
in some disciplines where difference-in-differences is seen as a more
credible design than simply controlling for observables. However, in all
the papers that propose this method, the approach is not rigorously laid out. In particular, while
there is testing for parallel trends in the main effects with a linear
model, there is no attempt made in the papers to make sure trends are parallel for
each CATE estimate. This is a problem because with machine learning of
heterogeneous effects we no longer have just the one estimate we need to
make this assumption for but a whole distribution of estimates. 

Presumably for inference over the conditional average treatment effect on the CATT we would have to make a version of a parallel trends assumption for within each kernel bandwidth. More formally,
where we target outcome $\Delta(W, X)\equiv Y(W,X)_{t}-Y(W, X)_{t-1}$ for the estimation of $\tau(x)$ via the adaptive kernel $\alpha(x)$ we must assume that $$\forall x, \quad \mathbb{E}[\alpha(x)\Delta_{t}(0,X_i) |X_i, W_i = 1] = \mathbb{E}[\alpha(x)\Delta_{t}(0,X_i) | X_i, W_i = 0].$$That is that in expectation the untreated potential outcomes are parallel for the weighted average of the datapoints in the neighbourhood of $x$. 
This is obviously a much stronger assumption than a parallel trends assumption at the average level and one that is harder to test, but this difficulty does not mean we can afford to just ignore the assumption. A proper treatment of
difference-in-differences and the causal forest is out of scope of this
paper, but these unresolved problems should make applied
researchers wary of naively using differenced outcomes as the target
variable in their causal forest.

One more unorthodox approach inspired by difference in differences is the one proposed by \citet{turjeman_when_2024}, the Temporal Causal Forest. They have a sample which was all treated at one time with a data-breach of website users, but the sample joined the website at different times, so they can counterfactually model normal fall-off of users by looking at the change in user activity before the data breach versus after the breach happens. Essentially it is a difference-in-differences across time since joining. This is an interesting approach at least for their specific research question and dataset. However, as with the more conventional first differences approaches it still requires an untested parallel trends assumption that may be violated for some neighbourhoods when calculating CATEs.

\subsection{Choice of variables}\label{choice-of-variables}

In addition it is worth speaking about the way in which covariates are
used. There is little distinction in the papers between the use of
variables in the nuisance models and the use of those variables in the
final stage model. Only seven papers clearly specify a set of
heterogeneity model variables that is different from the nuisance
function variables. Several others specify different variables in a
data-driven way, commonly this is with a method I will term the Basu Technique. The Basu Technique is a method for selecting variables for use in a model, it was first applied to the causal forest in \citet{athey_estimating_2019}. It fits a first model, takes the most important variables (measured according to methods discussed in Section \ref{subsec:varimp}) and then fits a final model on variables with only above-average importance. This approach aims to
decrease the amount of splitting that is just due to noise to provide greater clarity in results. Relatedly, it is a way of grappling with the calibration problem of having too many noisy features to be able to accurately estimate effects. 14 papers
employ this technique. Interestingly, they are not the studies with
larger sets of covariates. In fact the papers using the Basu Technique
had a median of 11 original variables compared to 18.5 for those that did not. There seems to be no clear guide as to
when it is appropriate or not appropriate to use the Basu Technique and
no methodological work on the improvements it may lead to in modelling
either whether in improving estimation or to make it easier to extract insights from a model. In particular there is no work to the best of my knowledge discussing the use of this technique in light of the drawbacks of commonly used variable importance techniques as discussed in Section \ref{subsec:varimp}.

\begin{figure}[h]
\centering
\includegraphics[width=.6\linewidth]{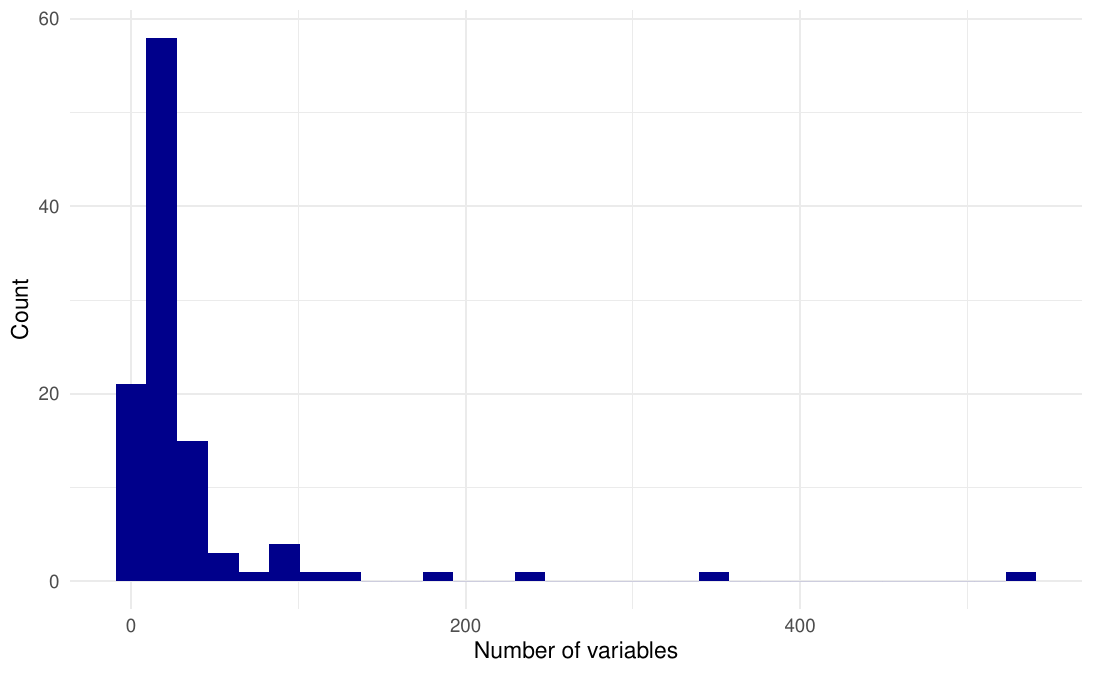}
\caption{Number of covariates variables used in each study for which the covariates are given (if separate variables are used across the three models e.g. if Basu Technique is applied, this is the highest count)}
\label{fig:figurebasu}
\end{figure}

The number of variables used is also generally small with the median
number being 17. Figure \ref{fig:figurebasu} shows the distribution of covariate counts.
While some designs use many variables, it seems obvious that for the
most part, the ability for the causal forest to excel with very
high-dimensional datasets \citep{gong_heterogeneous_2021} is not being
used. This may be because they lack access to sufficiently
high-dimensional data or because they do not see a large number of
variables as being useful; they might have an existing theoretical framework that narrows down the variable set. In the final model, it may also be that the
problem of extracting insights from CATEs calculated with a
large number of variables makes using large datasets more trouble than it is worth.

As an aside, the approach to sampling variables for use in the individual trees by default draws $\sqrt{p}+20$ variables in the grf causal forest. This means for most papers in our sample, the causal forest is not so much a kind of modified random forest, but instead a modified bagged trees ensemble where different trees are fit on different subsets of the data but the full set of variables \citep{breiman_random_2001, breiman_bagging_1996}. This is not necessarily a problem, but it is worth keeping in mind that the decorrelation of trees via random variable selection is the key insight of the random forest and when this parameter is left at its default value, we may be hurting the out-of-sample fit of the ensemble.

\subsection{Model specification choices}

Interestingly, because most approaches use the stock grf, no papers rely
on nuisance models that are not a random forest (though some use other kinds of models for matching before applying the causal forest and one uses a locally linear forest for propensity score estimation). This is not a bad
choice, particularly given the small size of the data in most studies
(both in width and length) and that grf is designed to easily fit these
nuisance models under the hood. However, it is certainly a choice worth
noting.

There is a little more diversity in the way random forests are tuned.
A number of papers papers used their own manual cross-validation of
hyperparameter values rather than relying on the grf tuning
forest \citep{guo_effect_2021, xu_using_2021, seitz_individualized_2023, verstraete_estimating_2023, inoue_heterogeneity_2023}. It is unclear from the text of those papers whether this
decision was made due to a mistrust of the tuning forest or whether the
authors simply did not know about this functionality and relied on a
more traditional ML fitting process instead.

The number of trees used varied widely from 300 up to 100,000 with a median of 4000 (for those that reported a number). 69
papers do not specify a number of trees used which suggests the default
hyperparameter value in grf of 2000 was used. The number of trees in the random
forest is an unusual hyperparameter in that unlike many others in
machine learning, tuning the number of trees should never
cause overfitting --- there is no trade-off to increasing the number apart from the computational cost \citep{breiman_random_2001}. In expectation, more trees will be
better than fewer and the main reason for choosing a smaller number is
simply runtime and memory needs. It is therefore better to err on the side of more trees
than fewer. It is after all relatively inexpensive to fit more trees and papers should simply grow the ensemble until there is essentially no excess error (the estimate of Monte Carlo error from having a finite-sized ensemble).

\subsection{Sample splitting}
In some cases authors used the standard machine learning approach of splitting their sample into a training dataset and a dataset for estimation (analogous to a test dataset) \citep{seitz_individualized_2023, kianmehr_machine_2022, serra-garcia_incentives_2023}. Because the causal forest always fits out-of-sample --- that is it only predicts for data-points based on trees that were not fit on that data-point --- this is redundant \citep{athey_generalized_2018}. In fact, in expectation, there will be no effect apart from shrinking sample size. The only reason this might be worth doing is if there is some sort of policy being derived from the forest for which one is estimating a value as estimating value in sample will probably overestimate due to over-fitting. A straight-forward example of where this is useful is \citet{jakobsen_machine_2023} or \citet{coleman_how_2023}, while \citet{osawa_targeted_2023} uses a more unusual approach without i.i.d. data between samples (validating on experimental data) which nonetheless is an interesting approach which could be useful for future policy learning work.

\section{Results - Presentation of
results}\label{results---presentation-of-results}

One serious challenge facing the causal forest is that even if there are
useful patterns in results, it is not obvious that researchers will be
able to identify and communicate these patterns \citep{rehill_transparency_2023}. The reason heterogeneity can be challenging to understand is that the
results are essentially a high-dimensional distribution of effects that
must be simplified in some way to allow for human comprehension. This
means that generally a causal forest needs to be explained with the use
of some kind of additional model. This could be a simpler model fit on locally centered data or doubly robust scores like
a best linear projection or a best causal tree (akin to interpretable AI
in predictive modelling \citep{rudin_stop_2019}). Alternatively, one can look at the predictions
made by the model and try to explain them like through binned CATEs or a Rank-Weighted Average Treatment Effect (RATE) \citep{yadlowsky_evaluating_2023}.
Finally, one can look at variable importance which tries to reveal
relationships by analysing the structure of the tree itself. The focus
of this section is not on all the possible approaches but instead on discussing the
practices in this sample where there are a few that are used widely.

There are three common ways of visualising results, most papers use one
or more of these. Graphing of the overall
distribution of individual treatment effects (individual CATEs were graphed in a histogram in 33 papers and a ranked plot of treatment effects was used to a similar end in 9),
graphing of conditional treatments effects often with binning for
continuous variables or a kernel regression (used in 58 papers) and
showing variable importance (used in 35 papers). Table \ref{tab:table2} shows how common these
and other approaches are. Several other approaches were used only once. These are presented in an expanded version of this table in the Appendix. The rest of the section will run through some of the most popular ways of presenting results.

\begin{table}[]
\begin{tabular}{ll}
Method                                                                    & Count \\
Graph across key variables                                                & 58\\
Variable importance                                                       & 35\\
Graph individual CATEs in a histogram& 33\\
Graph along quantiles of CATE estimates                                                     & 26\\
Single interpretable tree                                                               & 15    \\
Best Linear Projection& 12\\
Derive policy& 11\\
Rank Average Treatment Effects (RATE) \citep{yadlowsky_evaluating_2023} or similar& 9     \\
SHAP                                                                      & 6     \\
Fit linear model on causal forest predictions& 7\\
Graph across geography& 5\\
Fit quantiles for individual CATEs then look at covariate distribution within quantile& 5     \\
Partial dependence plot                                                   & 3     \\
K-means clustering                                                        & 2    \end{tabular}
\label{tab:table2}
\caption{Number of times a particular technique was used in the studied papers to communicate CATE results.\\
Note: A detailed explanation and table included approaches only used in one paper is presented in the Appendix.}
\end{table}

\subsection{Graphing individual treatment effects}
A univariate plot graphing out the distribution of the individual CATEs can give some sense of how
much heterogeneity is in the data. While this is not always the case, it
can be a good basic check on data and in some cases as in \citet{guo_effect_2021} it is very expressive, laying out a pattern of treatment
effects that is very useful in answering the research question. However the basic approach of graphing a histogram of predictions is not particularly useful. It is impossible to separate out variation here from noise in estimates around a constant null effect. Graphing
of ranked effects is a variation on this which shows more
information because it allows for easy visualisation of errors along with point estimates. While arguably the point estimates are a little harder to interpret here, such a plot does allow for visualisation
of confidence intervals on the same plot ala \citet{fukai_describing_2021}.
Another variation on this is the graphing of CATEs across quantiles of
CATE estimates. Here estimates for each treatment effect quantile are made using held-out data because otherwise the partitioning and estimation will not be independent, leading to a kind of overfitting where the lower quantiles will have misleadingly low estimates and the high quantiles will be misleadingly high. One variation on this approach is that from \citet{davis_rethinking_2020} where the top quantile is compared to the other quantiles in order to get a heuristic of the amount of heterogeneity in a distribution \citep{mesple-somps_role_2023, davis_rethinking_2020}. Two papers actually statistically tested the difference in quantile effects with a Wald test \citep{serra-garcia_incentives_2023} or a Kruskai-Wallis test \citep{yu_smart_2024}.

\subsection{Variable importance} \label{subsec:varimp}
Variable importance is a simple metric but a useful one, and one that is
easy to explain to a reader. It takes a depth weighted count of the splits on each variable for all trees in the forest (by default halving the weight of each layer compared to that above it before stopping counting at depth 4) and the normalises there to sum to 1 \citep{tibshirani_package_2021}. Many papers either present variable importance values directly or use it to drive their search
for treatment effect heterogeneity where they lack a good theoretical
framework for which variables to use. It can also help to check the
robustness of the theoretical framework in the particular dataset being
studied. That said it is certainly just a heuristic and is not as robust
as techniques for variable importance in predictive modelling. These predictive metrics can better partial out
the effects of variables on estimates because they have access to ground truth \citep{strobl_conditional_2008}. It is worth noting that the mcf Python package which implements the modified causal forest \citep{lechner_modified_2019} does implement a permutation variable importance. No papers here use this approach. 

Variable importance becomes less useful in cases
where there are many variables because highly correlated features \citep{benard_variable_2023}. This is because correlated features effectively split importance between them. Newer
approaches propose more robust variable importance metrics, but at the
cost of longer runtime and not having convenient implementations already
available \citep{benard_variable_2023, hines_variable_2023}. Only two papers use an approach other than that from grf for variable importance \citep{venkatasubramaniam_comparison_2023} uses a permutation test which seems similar to that proposed by \citet{benard_variable_2023}. In theory this permutation test is likely to be superior although can become massively computationally difficult in even moderately sized datasets thanks to the algorithm being combinatorially complex. It is also of course unclear to what extent such a complex operation is necessary given that unlike in predictive models, we cannot actually observe the effect of each variable on predicting ground-truth. It is possible to imagine this might be a powerful approach for analysis with just a handful of variables though for the reasons laid out by \citet{benard_variable_2023}. \citet{dylong_biased_2024} use their own approach that is slightly different, they look at the difference in estimates for each quartile of a covariate which they call relevance. It hard to assess the usefulness of this relevance approach given it is only used in one paper. This may be a useful approach but it may also fall victim to noisy features in the way \citet{benard_variable_2023} fears standard variable importance does, as noisy features are also more likely to vary across quantiles. Further work to assess the value of this approach is needed.

\subsection{Estimating CATEs across variable values}

\begin{figure}
    \centering
    \includegraphics[width=0.75\linewidth]{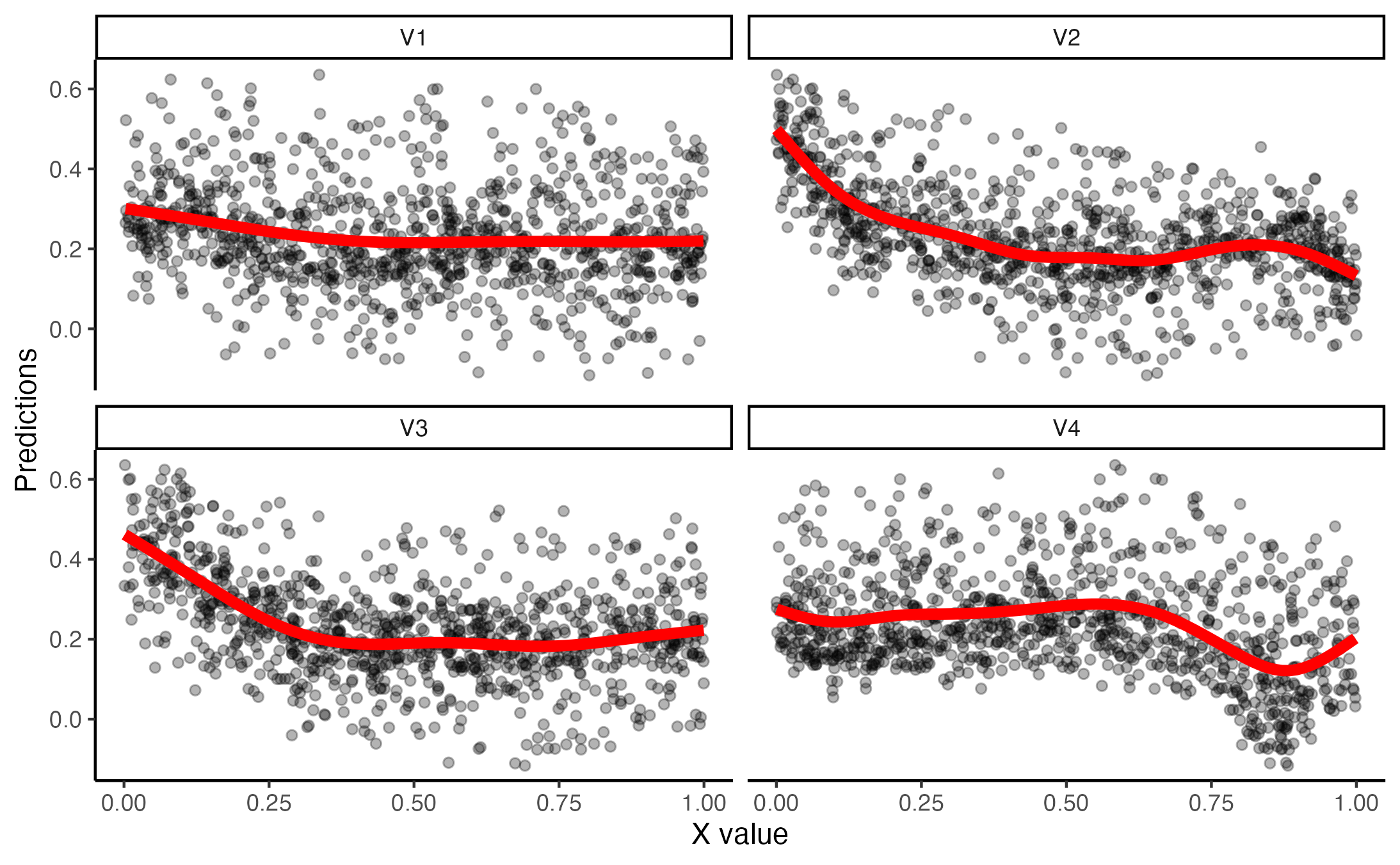}
    \caption{Example of graphing predictions across covariate values including a smoothed conditional mean with a lowess regression line.}
    \label{fig:eg_across}
\end{figure}

Graphing effects across variables can be done in a number of ways. There are many ways to choose which variables to estimate across (based on variable importance or based
on theory) and they can be shown in many different ways (e.g. boxplots
for binned or binary variables, scatter plots with a kernel regression
line fit on predictions for continuous variables, or simply reporting subgroup estimates in text). One example is shown in Figure \ref{fig:eg_across} which shows predictions for a simple four variable data generating process with data generated according to the second DGP in \citet{wager_estimation_2018}. While the
standard approach was simply to graph CATEs as
a function of a variable, there were many variations on this approach.
Some papers used partial dependency plots brought across
from the explainable AI (XAI) literature \citep{seitz_individualized_2023, xu_principled_2023, esterzon_enhancing_2023} which if valid offer a better
view of the conditional effect of a given variable. In practice this may not be
appropriate given the nature of the causal forest model not itself
making predictions but instead fitting a kernel for use in an
estimator \citep{sverdrup_add_2021}. A variation on this is the use of SHAP, a
similar XAI method that is extremely expressive but again suffers from
the same concerns about ignoring the model's purpose as an adaptive kernel. This is not to say that neither approach is
appropriate, but rather there is no good theoretical work so far that establishes these XAI approaches are appropriate \citep{sverdrup_add_2021} (though it is worth
noting that SHAP is built in to the EconML Python package \citep{oprescu_econml_2019} so is likely
being widely used in industry applications).

An approach appropriate for some research questions was to graph CATEs for different areas on a map \citep{kluger_combining_2022, yin_effects_2023, deines_recent_2023}. This seems to be quite a
clear and expressive way to show treatment effect variation. 

When estimating CATEs across pre-specified variables is taken to its extreme, the causal forest becomes superfluous \citeg{esposti_non-monetary_2024}. Essentially this is just a series of doubly robust ATE calculations for different subsamples for example different age brackets. This is easy without the causal forest. At most grf faciliates the easy estimation of doubly robust scores.

\subsection{Deriving a single tree}
15 papers derived a single expressive tree from the causal forest which
they used in some way to communicate results. This is potentially a
useful method which could capture interaction effects in CATEs in a way
other ways of exploring CATEs do not. However, the usefulness of the
tree depends on how it was derived. There is no clear best practice to
derive a tree but there are many possible options whether a tree is selected based on some criterion from the forest \citep{wager_find_2018} or simply fit on the whole dataset separately. Interestingly, no
paper seems to use the R-Loss based selection method recommended by \citet{wager_find_2018}. Instead it is common for a single tree to be fit independently or
selected based on a tree-distance metric \citep{zhou_targeting_2023, zhou_heterogeneous_2023}.
The aim of the tree distance metric is to select a central tree in an
ensemble \citep{banerjee_identifying_2012}. This is certainly a valid approach to
the problem of selecting a single tree where we lack good ground-truth
data. It is hard to say which is better, this approach or minimising
R-Loss \citep{wager_find_2018}. While we obviously want a tree that performs well and so might
reduce R-Loss, it is also a very noisy objective so whether this
explicit optimisation actually gives a more `representative' tree is not
entirely clear. This question depends on what we actually want from a
single tree (representativeness or fit) and how much we trust the
intuition of Wager over these three papers. Finally it is possible to simply take a tree at random as representative (this seems to be what \citet{amann_effect_2023} does).

When fitting a separate tree it is possible to fit a tree separately on data as many papers do \citeg{giannarakis_towards_2022, rana_machine_2019}. However, it is also possible to distill a tree from the causal forest which may improve its stability and fit \citep{frosst_distilling_2017}. In fact the properties of the causal forest which give it good properties in an estimator (based on the insights of the semi-parametric estimation literature) may make it particularly amenable to distillation too \citep{dao_knowledge_2021}. \citet{rehill_distilling_2024} recently proposed an approach for doing this for the causal forest. That paper shows distillation generally outperforming alternative approaches listed in this review, particularly in noisy data-generating processes.

\subsection{Studying a policy derived from CATE results}
Several papers also look at explicit targeting of treatments whether
this is based on a `black box' policy from the causal forest or a second
interpretable policy learning model fit on the forest. In the first
case, targeting is based directly on the CATE estimate such that for policy function $\pi(x)$ a simple black box policy would be $$\pi(x) =
\begin{cases} 
1 & \tau(x) \geq 0, \\
0 & \tau(x) < 0.
\end{cases}$$The threshold may be set at some other value, for example at the cost of treatment or at a value determined by a budget constraint. A more interpretable approach to learning a policy can be to fit a new interpretable model on a set of variables that are useful for setting eligibility. One approach for this is to learn a model as a weighted classification of estimated benefits \citep{wager_causal_2024}, for example in the policy tree approach developed by the same team responsible for grf \citep{athey_policy_2020, zhou_offline_2023}. An interpretable approach of course has the benefit of being interpretable in trying to understand effects, however if decision-makers might be interested in using analysis to actually determine policy, an interpretable policy can be much more practical \citep{wager_causal_2024}. There are certain characteristics that might not make sense as bases of setting eligibility and others which may. For example, valid factors for discriminating on treatment may need to meet social ideas of justice, be difficult to manipulate, and be easily measured in the population seeking the treatment. In general policies were used as a way to understand heterogeneity, there was little discussion of actually implementing policies. 

Papers commonly followed the black box approach (11 papers did so), which is interesting given the policytree package in R interfaces well with grf; sharing many of the same authors \citep{sverdrup_policytree_2020}. The only paper to explicitly fit a policy tree was \citet{cockx_priority_2023} which also looked
at the effects of the `black box' policy as well. Perhaps the policy
tree is a poor fit for empirical researchers and may be more used in
other types of work where direct prescriptions are more relevant.

The black box policy analysis seems to have been useful particular where
research is designed to directly inform practice (many of these papers
were around changing treatment regimes in medicine). It gives a sense of
the potential gains of incorporating the knowledge from the study. The
use of a held-out test sample ala \citet{osawa_targeted_2023} makes sense here
in a way it does not for other applications of the causal forest because
the policy is in effect set based on the CATE estimates. We would expect
the estimate of the policy value to overfit in-sample even though the
estimates themselves are unbiased as causal effects. The way this learning is visualised varies too, but curves
comparing outcomes under the `treatment' of policy targeting and the
standard regime like in \citet{osawa_targeted_2023} or the Qini curve used in \citet{seitz_individualized_2023} are good examples of how this information can be
communicated. There are likely to be more insights on this that can be
drawn from uplift modelling which tackles similar problems \citep{zhang_unified_2022}. 

Another version of this is \citet{jakobsen_machine_2023} which uses a RATE analysis. This is part of a fairly extensive effort to uncover treatment effect heterogeneity. In this case, the treatment is not really manipulable (COVID infection) so the direct utility of a policy is low, but the authors seem to find it useful for presenting their forest results.

\subsection{Use of linear models to explore heterogeneity}
12 papers relied on fitting a best linear projection (BLP) \citep{semenova_debiased_2021} to explain
heterogeneous effects \citep{wang_effect_2022, zhang_inferring_2022, elamin_overeducation_2023, huber_business_2022, esterzon_enhancing_2023, xu_using_2021, elek_regional_2021}. A best linear projection regresses doubly robust scores onto a set of variables in a linear model. In grf, standard errors are calculated via resampling to adjust for having a dependent variable that is itself estimated from a model. While there are obviously limitations here (it
can only measure linear effects where the rest of the approaches have been
free of functional form limitations), it could be a useful way to
hypothesis test results to get an output that is more easily
interpretable to an audience more familiar with parametric regression
outputs. Generally the BLP was used to projects the effects onto the same
variables that were used to fit the causal forest in the first place.

A number of studies tried something similar but did not project effects
using the approach built-in to the grf package, they instead fit a linear model on CATE estimates directly directly as proposed by
\citet{nilsson_assessing_2019} \citep{habel_effective_2023, leite_heterogeneity_2022, guo_effect_2021, chen_improving_2020, zhu_living_2022, amann_effect_2023}. This is a bad idea for two main
reasons. The first is that fitting a linear model on predictions rather
than doubly robust scores means the model lacks double-robustness and is more likely to be
biased. It is essentially a conditional average of a series of
conditional averages (the predictions) the offer no guarantees about the
properties of this conditional mean. The second and I would argue worse problem is that it seems like in all
these cases, the standard errors are calculated analytically as is
standard for linear regression rather than with resampling. This does not meet the assumptions for valid
analytical standard errors because doubly robust scores are outputs from another model \citep{wooldridge_introductory_2013}. In order to get valid
standard errors in this approach with looser assumptions, we need to
estimate errors through resampling across all models like the grf BLP implementation does automatically. Papers that fit their own
models on predictions do not seem to do this. The result is a biased point estimate and likely over-confident
standard errors. When fitting a model researchers should always use the
procedure built into grf rather than attempting their own.

\subsection{Using a clustering model}
Finally, two papers clustered treatment effects with an unsupervised
learning algorithm (K-means or K-means++) \citep{miao_effects_2023, cockx_priority_2023}. This could be a promising approach as it accounts for
covariation between variables in a way that clustering based on a
single tree does not. Hopefully this approach will be developed further
in future papers such that its value across a wider domain of problems
can be assessed.

\section{Discussion}
\subsection{What has been done so far}
The causal forest
literature is currently largely defined by an implicit best practice set
out by the creators of the method, either in the way they have used the
method in their own work (such as the use of the Basu Technique for
dimensionality reduction) or the way in which they wrote the grf package
(such as in relying on default hyperparameter settings). This is not a
bad thing at all --- most of the criticisms I have made in this paper have
been of papers stepping outside this practice --- but it does mean there
is for better or worse a methodological conservatism,
at least in the applied space. As a side note, there are many methods
papers that propose novel modifications to the causal forest, however
these have yet to be adopted in the applied literature so far.

The review finds that the causal forest has been used in many different ways across many different
fields, however, most publications stick to the
same basic design laid out by Athey, Wager and co-authors across several
different papers \citep{wager_estimation_2018,athey_generalized_2018,athey_estimating_2019, athey_recursive_2016}. This approach has been encouraged by the
excellent tooling in their widely used grf package. This
approach generally involves fitting a causal forest via grf on observational or
experimental data using somewhere around 2000-4000 trees and 10-20 covariates for each model. Often users trim the number of variables used in the final model
by using the Basu Technique laid out by \citet{basu_iterative_2018} and demonstrated for
the causal forest by \citet{athey_estimating_2019}. Rarely, a researcher specified a different set of nuisance and heterogeneity model variables, however for the most part the same features were used for each. This is interesting because while the data-driven selection of important variables via machine learning in some ways makes it less important to manually select features, it also suggests that researchers may not be taking the role of moderators seriously or lack data on moderators. Moderators of course are not necessarily pre-treatment variables so may not be valid variables for inclusion in the nuisance models but may be important predictors of heterogeneity.

Identification is generally achieved either
by randomisation or through controlling on observables. An
ATE is then estimated using augmented inverse propensity
weighting (AIPW) \citep{robins_estimation_1994}. 

CATEs are explored by
some combination of plotting the distribution of individual estimates as
a histogram, presenting variable importance (that is a depth-weighted
count of the use of different variables in splitting), fitting a best
linear projection \citep{semenova_debiased_2021, tibshirani_package_2021}, or graphing treatment effects across values of important variables.

Often (particularly on designs with observational data), this causal
forest analysis is one part of a paper that uses more standard methods
to estimate an average effect before using a causal forest to explore
treatment effect heterogeneity or as a robustness check for their preferred method of studying heterogeneity. For example, it is not uncommon for a paper to use a traditional, parametric difference-in-differences approach to estimate an average treatment effect for the treated and then use a control-on-observables causal forest to understand heterogeneity.

While there is a lot of similarity in approaches, some authors are innovating; taking on methods developments from outside of the Athey and
Wager body of work \citeg{cockx_priority_2023, osawa_targeted_2023} or presenting results in
innovative ways \citeg{baum_targeting_2017}. However, not all these novel approaches are
positive. Some changes seem to be redundant or even harmful. For example, several authors use cross-validation for hyperparameter tuning when there is already a tuning forest built into the grf causal forest that does tuning automatically \citep{guo_effect_2021}. In all of these cases, there are
reasons why the authors may have done this, but the papers do not make a
point of justifying this approach. This suggests it is simply due to a
misunderstanding of the causal forest implementation they are using.
There are also cases where authors seem to misunderstand the method and
do things that are outright counter-productive such as averaging
together individual tree predictions rather than using the forest to
provide an adaptive kernel for an estimator \citep{bittencourt_evaluating_2020} (thereby using the older method from \citet{wager_estimation_2018} and ignoring
the generalised random forest \citep{athey_generalized_2018}).

While this paper has been skeptical about some of the attempts authors
have made to innovate on causal forest methods outside of the standard
grf best practice, sensible innovations that are well-justified and
build on a strong knowledge of the existing methods could offer
substantial benefits for creating more flexible and more transparent
methods in the future. Certainly on the former, new approaches to
identification with the causal forest could dramatically improve the
credibility of the methods in fields such as applied economics where
control-on-observables is generally not viewed as a credible
identification strategy and where experimental data is often scarce. For
example, the work on difference-in-differences with the causal forest or
the use of causal forest with regression discontinuity are areas that
could be explored much more than they have been. As a first step, with a sufficiently large dataset it would be trivial to use a causal forest within a local randomisation window where treatment allocation can be treated as as good as random \citep{cattaneo_randomization_2015}. A more involved approach would be building identification into a causal forest algorithm, for example by incorporating parallel trends into the objective function of a difference-in-differences causal forest by including a penalty term based on \citet{rambachan_more_2023} error bounds for parallel trends violations. In addition new
approaches to estimation can open up new research questions that simply
cannot be answered with the standard grf approach. For example, \citet{cockx_priority_2023} studying policies derived from the causal forest is a good
way to understand how the forest can add value in shaping actual policy
interventions. \citet{osawa_targeted_2023} testing the policy on a second
validation sample of experimental data is going to be useful in some
contexts to test whether the value of the policy estimated in the policy
learning step generalises to new data.

On transparency, methods for understanding effects are still very much
lacking, posing problems both for researchers who are trying to extract
insights from their analysis and also, for those whose lives might be
affected by decisions made on the basis of causal forest analysis
\citep{rehill_transparency_2023}. One promising area here is in more robust
methods for calculating variable importance \citep{benard_variable_2023, hines_variable_2023}, more akin to the way variable importance is calculated in
predictive models than the heuristic implemented by \citet{tibshirani_package_2021}. These promise at the very least to be able to distinguish between important splits (those that change the CATE distribution greatly) and unimportant splits (those that have minimal effect) and be able to account for correlated variables through a process of fitting different models with different sets of variables. Of course, the trade-off here is the great computational complexity of fitting many different forests with many different sets of variables. Another approach that
could be valuable is to use SHAP values as the EconML package allows,
however the theoretical work has not been done yet to establish this as
a valid approach.

The papers in the study are all peer reviewed, applied research
papers written largely by academic researchers, this has its benefits but
it is also a limitation. There is a whole other area of activity where
the causal forest and similar methods are being applied to
solve problems in industry, often called uplift modelling problems in
this context. It is also worth noting that by only looking at peer
reviewed papers, in a relatively new and fast-moving field this paper
may have missed important advances that have so far only been made in
papers published as preprints. Finally, there are many other causal machine learning methods where estimates are made differently but which have some of the same challenges as the causal forest, for example Bayesian causal forest \citep{hahn_bayesian_2019} (which despite its name is a very different method from the causal forest) or causal meta-learners \citep{kunzel_metalearners_2019}. \citet{chernozhukov_fisher-schultz_2023} already has some interesting ideas (aside from the BLP) that could be more widely used with causal forest CATE distributions. There may be lessons to be learnt from this broader literature as well. 

One potentially troubling factor in this sample is that many papers have quite low sample sizes. While the random forest performs well with smaller datasets compared to other machine learning approaches \citep{athey_generalized_2018, breiman_random_2001}, it is still not clear that data-driven HTE analysis is appropriate in these settings. For starters there is no real understanding of statistical power for kernel bandwidths in the causal forest. CATE estimates that use a small part of an already small sample are likely to be prone to type II error. Worse, the way the causal forest is often used does not just risk type II error in the way a parametric model might. Because the forest is effectively generating hypotheses, helping a researcher to understand the kinds of things they might want to test, it can be even more problematic. Many of the commonly used outputs like graphs along key variables and variable importance give no confidence intervals and so a researcher's understanding could be warped without their knowing. If they then go on to run tests based on this knowledge for example in picking subgroups for CATE estimation or in choosing projection variables for the BLP they have tailored their tests to potentially spurious effects in the data. One solution to this might be to run an global test for heterogeneity first before moving to detailed CATE analysis for example via a RATE estimation on held out data \citep{yadlowsky_evaluating_2023} or grf's built-in calibration test.

There is one last point worth making which relates to the task of assembling a review but also broader reproducibility and interpretability of research. Many researchers did not specify basic information about their use of the causal forest. What packages they used, the number of trees they used and the like. Unlike some more standard methods, the causal forest has several possible implementations all called the causal forest, it has hyperparameters that need to be picked and it is nondeterministic. There are a lot of choices that need to be made when modelling and a lot of information that should therefore be given in papers. Providing code is especially useful here. While it is always useful for researchers to make their code
publicly accessible, the process of compiling this review has shown just
how important that can be when using a causal forest. While some authors wrote detailed (and much appreciated)
appendices on their specification, access to code is invaluable for
understanding the exact approach a paper took. In many cases, the
precise specification of a model was unclear from simply reading the
published paper.

\subsection{Reflections for future practice}
This section will briefly pull out some high-level advice on the use of causal forest methods in research based on the good and bad practice discussed in this paper so far. 

\subsubsection{Data needs}
The causal forest should be applied in cases where there is a reasonable number of observations, though there does not currently exist a good rule of thumb as to how large this is. One of the advantages of using a random forest-based model is that the approach can perform well on what are small datasets by the standards of machine learning practice while still performing well in datasets that have too many variables for more traditional methods of studying heterogeneity (like a linear model with interaction terms).

While more variables generally means more accurate estimation of the CATE, there is no reason that a causal forest should not be used where only a few covariates are available (as in many experimental datasets). The actual accuracy of the estimates is less useful than the ability to trace the effects of different variables (at least when using the forest just for research purposes, as opposed to prescriptive purposes where overall accuracy is more important \citep{rehill_transparency_2023}). On the other hand, a large number of variables, particularly where there are few useful predictors hidden amongst a lot of useless ones can hurt performance \citep{fiagbe2021}. This is because when a subset of variables are made in each tree, there is a low probability that useful features are in the dataset and so splitting becomes dominated by noise. Some kind of variable selection mechanism could be useful here but this is an open problem, some solutions include selecting based on variable importance \citep{basu_iterative_2018}, or using knowledge distillation to remove noise \citep{rehill_distilling_2024}.

\subsubsection{Identification}
Section \ref{sec:id-strat} already touched on the different identification challenges here a little. To summarise, if there is enough data available, an RCT is the ideal case using the oracle propensity scores (experimentally assigned treatment probability) in estimation rather than allowing the causal forest to estimate its own propensity score nuisance model by manually setting the \textit{W.hat} argument in \textit{causal\_forest}.

Quasi-experimental methods are much less used in practice and there has not been much methodological development around these methods, however there are techniques for difference-in-differences, regression discontinuity and instrumental variables designs. Building up the tooling for these further is an important open research question.

On control-on-observables designs, while simulation studies on double machine learning, R-Learner and the causal forest can be helpful, ultimately there will never be a perfect answer here and choices about design will have to be shaped by disciplinary norms and knowledge about the data-generating process. If these are not taken seriously, causal machine learning approaches could do more harm than good, encouraging sloppy designs that are deemed acceptable because of baseless confidence in the power of machine learning models to model out confounding \citep{rehill_fairness_2023}. 

In an example of a good design, causal machine learning approaches using control-on-observables designs have been shown to be useful when studying electronic health records as these can be processed as high-dimensional data and almost all of the information that doctors will use to make treatment assignment judgments will be encoded in a patient's health records \citep{ross_estimated_2024}. In addition, health fields reliant on observational data like epidemiology generally take the task of modelling out confounders very seriously with tools like target trial emulation \citep{hernan_target_2022}. On the other hand, the norms and kinds of research projects tackled in econometrics might mean a control-on-observables design is less appropriate here. For example, in a policy evaluation of a job training program where there is scant pre-treatment data on participants and there is no measure of latent psychological confounders, machine learning won't magically allow for identification. In addition, the norms of econometrics will likely treat such a design very skeptically. 

Finally, in order to satisfy ignorability, overlap needs to be taken seriously. This was not the case for most control-on-observables papers in this sample which did not explicitly test overlap or adjust their sample to remove regions of poor overlap. As already mentioned, in an ideal world we would know that this assumption holds in every bandwidth used in estimation, but at the very least, validating this assumption at the overall average level should be a basic step in any control-on-observables design.

\subsubsection{Estimation of effects}

Software like grf makes estimating CATEs relatively simple. However, there are still a few potential challenges that it is worth being aware of during the fitting of a model.

The first is around the dimensionality of data used. If a causal forest is fit on a very large, sparse, dataset this can lead to a poorly calibrated model that is a poor estimator. How one should go about dealing with this problem is something of an open question. The most widely used approach as already mentioned is the Basu Technique. However, this has its drawbacks. In the same kinds of settings where it may be useful to do some kind of dimensionality reduction like this, variable importance is likely to become unreliable. Further work is needed on whether it is likely to be so unreliable as to be useless. In addition, there is a broader literature on overcoming a large number of non-informative features when using the random forest which could inform future causal forest approaches \citep{fiagbe2021}. This is not to mention the possibility of using traditional dimensionality reduction approaches (though there are obvious problems with using many of these out-of-the-box like linearity assumptions or difficulty interpreting effects on pre-processed data).

The second issue is around testing for heterogeneity in estimates before digging into specific patterns. It can be useful to know before proceeding any further if there is actually sizable heterogeneity in the data. There is no single established way to do this. The easiest may be to look at the result of the test\_calibration function in grf. This function is intended to measure how well the causal forest fits variation in treatment effects but can also be a useful test for heterogeneity in data. If the coefficient for the predictions of the causal forest is significantly different from zero, that is a good sign there is significant heterogeneity in the data. Another approach to testing for overall heterogeneity would be to fit a RATE curve on held-out data to see if there if a significant difference in estimated treatment effects across treatment prioritisation \citep{yadlowsky_evaluating_2023}. A more in depth discussion of an overall test for heterogeneity can be found in \citet{sverdrup_estimating_2024}.

\subsubsection{Presenting estimates}

When estimating CATEs there are many different ways to present findings. A rundown of all the ones identified in this paper can be found in the Appendix. Not all of these approaches are necessarily valid. The use of a linear model fit on predictions out of a causal forest for example is used several times in the papers studied but is not statistically valid for inference. The issues around some approaches have already been covered in Section \ref{results---presentation-of-results}. 

Here are a few high level reflections on the most common methods. Graphing effects across important variables can be useful, the key is how the important variables are chosen. If they are chosen based on ex ante theory, be aware that you are choosing for better or worse to ignore what the causal forest thinks is important. If using something like variable importance instead, there are drawbacks here too \citep{benard_variable_2023}. Variable importance, whether used for picking which variables to show or just presented as an output itself has its own drawbacks. In the causal forest setting, generally a computationally cheap heuristic method is used which can be unreliable, particularly with a lot of correlated features or noise in the dataset. The technique of graphing treatment effects in a histogram is of limited use, it takes extreme heterogeneity to show up something in this distribution. Usually it is hard to tell informative heterogeneity from noise in underlying treatment effects, and either of these from variance in the estimates. Graphing across quantiles of effect may be useful but it is important to have a sample split here where part of the data is used to allocate cases to each quantile and held-out data is then used to estimate a CATE for each quantile. This is because otherwise we would expect high quantiles to have effects biased upwards and low quantiles to have effects biased downwards (because the same data is used to allocate quantiles and estimate effects within them). Single tree approaches can be useful, particularly in high-dimensional settings. There are many different ways to fit a single tree, some are more robust than others \citep{rehill_distilling_2024}. A best linear projection is a good way of taking some of the findings of a causal forest study and communicating them in a way that is easy to understand for audiences who are not familiar with these methods, but are familiar with linear regression. Most importantly, this allows for easy inference around estimates. However, it relies on linearity of effect and does not deal well with high-dimensional datasets, so another method of selecting variables and testing linearity is important as well.

With so many options here, the choice of approach depends a lot on the specifics of the problem. In a setting with good ex ante theory, graphing effects along these important variables as well as some that have high variable importance is useful. In a high-dimensional problem where it could be hard to know which variables are important drivers of heterogeneity, something like a simplification to a single tree could be useful. In a study of effects across different regions, showing effects on a map could be powerful. In a study looking at CATEs for the purposes of introducing eligibility rules or prioritising cases, something like a RATE plot or derivation of a treatment policy may be useful to explore the prescriptive implications of the analysis.

One final point on this though, a balance needs to be struck between hewing to established practice and methodological innovation. This applies mostly to this area of presenting estimates but is applicable in some way to all the stages discussed here. The causal forest is a relatively new method, there is still a lot of work around building out the use of the method. New ways of using the causal forest can be very valuable within a paper, but also can add value for other researchers who can pick up the method in future. The problem here is that innovating around the use of a method like the causal forest can go badly if it is not treated with the seriousness it deserves. Bad methods can not only lead to invalid conclusions in one study, but the use of the method can be picked up in other studies as well leading to a harmful contagion effect.

\section{Conclusion}
The causal forest is a promising method that is being used more and more over time. This paper has sought to understand how this still nascent method is being used in order to reflect on and maybe even shape the emerging best-practice. The review shows the causal forest is largely used for research that achieves causal identification with randomisation or by controlling on observables, there is little in the way of quasi-experimental work using the method. Furthermore, the vast majority of work uses the grf package in R. There are a reasonable number of commonly used methods for presenting results and and a much larger set of methods that are used only rarely but which show --- for better and for worse --- that there is still some innovation happening in figuring out how to communicate insights from a CATE distribution.

No doubt the method will continue to evolve, but there appears to be a few key areas ripe for further work. Control-on-observables designs should do more to test for identification (for example assessing propensity score overlap which only a minority of papers did). Quasi-experimental variants of the causal forest may help to make identification more credible in some research. There is still a lot of work that could be done to improve the presentation of results, particularly useful would be the development of theoretically valid XAI approaches that account for the unusual process of getting causal forest estimates. Finally, it will be interesting to see whether larger datasets both in width and length begin to be used as researchers begin to embrace models that can better cope with these (and as they begin to get access to bigger datasets \citep{connelly_role_2016}). The promise of better being able to understand treatment heterogeneity through causal machine learning is an exciting one, and many of the papers reviewed here show this promise beginning to be realised.

\bibliographystyle{agsm}  
\bibliography{references}  

\newpage\section*{Appendix - Full table of presentation methods and explanations}

\begin{table}[!h]
\begin{tabular}{ll}
Method                                                                    & Count \\
Graph across key variables                                                & 58\\
Variable importance                                                       & 35\\
Graph ITEs in a histogram                                                 & 33\\
Graph along quantiles                                                     & 26\\
Single tree                                                               & 15    \\
Best Linear Projection& 12\\
Derive policy& 11\\
Rank Average Treatment Effects (RATE) \citep{yadlowsky_evaluating_2023} or similar & 9     \\
SHAP                                                                      & 6     \\
Fit linear model on causal forest predictions& 7\\
Graph across geography& 5\\
Fit quantiles for ITE then look at covariate distribution within quantile & 5     \\
Partial dependence plot                                                   & 3     \\
K-means clustering                                                        & 2    \\
GAM model based on important variables & 1\\
Graph predicted vs observed effects in future & 1\\
Graph probability of negative outcome & 1\\
PCA on most important variables & 1\\ 
Plot across variable holding others fixed at median & 1\\
Q-Score metric & 1 \\
Qini curve & 1\\
Slope test                                                        & 1 
\\
 Rank-weighted Average Treatment Effect&1\\\end{tabular}
\label{tab:tableapp}
\caption{Number of times a particular technique was used in the studied papers to communicate CATE results}
\end{table}

\textbf{Graph across key variables} --- graphs individual predictions of CATEs along values of a variable.

\textbf{Variable importance} --- presents some measure of how important a variable is, either by how often it is used or how much it affects estimates.

\textbf{Graph ITEs in a histogram} --- presents a histogram of the distribution of individual effects.

\textbf{Graph along quantiles} --- this graphs CATEs in some way (for example box plots) along quantiles of the CATE value. Most commonly just a median split.

\textbf{Single tree} --- takes a single tree in some way, whether this is by selecting a tree from the ensemble or fitting a new tree.

\textbf{Best Linear Projection} --- takes the doubly robust scores calculated with the nuisance model and makes a linear projection of them onto a selection of variables which may or may not be the same variables used to fit the forest.

\textbf{Derive policy} --- seeks to explain effect by fitting a series of rules to decide who should or shouldn't receive the treatment in order to maximise utility.

\textbf{Rank ITEs} --- similar to graphing a histogram. This approach takes the individual-level estimates and rank orders them to give a picture of the distribution of estimates.

\textbf{SHAP} --- an explainable AI method that gives local explanations of individual estimates using a game-theory based algorithm to assign different amounts of deviation from the ATE to different variables.

\textbf{Fit linear model on causal forest predictions} --- a statistically flawed method to make interpretable and hypothesis test causal forest estimates by using them as a dependent variable in a linear regression. Much like a best linear projection except this approach lacks the robustness and valid inference of that approach. 

\textbf{Graph across geography} --- where geography is an important factor in the study, this method visualises CATEs for each area.

\textbf{Fit quantiles for ITE then look at covariate distribution within quantile} --- this approach splits up the sample by quantile of estimates, but it then explores how covariate values differ across these quantiles. Variables that vary across the quantiles may be important drivers of heterogeneity.

\textbf{Partial dependence plot} --- this is an approach related to graphing across key variables except instead of graphing across predictions on observed variables. While it is not used much in this work, it has a longer history in transparency work for predictive machine learning. It takes each variable at a time, holds every other variable at its mean or median and varies that one variable across its range making predictions. Compared to graphing across the distribution of key variables, this in theory seperates out the effect of each variable from others that may be correlates.

\textbf{K-means clustering} --- this approach uses a popular clustering algorithm to find clusters in covariates and predictions and then analyse the positions of the cluster centroids.

\textbf{GAM model based on important variables} --- a GAM is a generalised additive model. This decomposes the predictions across key variables into an additive model of smooth functions. In practice this is not particularly different from graphing effect across key variables.

\textbf{Graph predicted vs observed effects in future} --- this approach can be used where new units will be treated and outcomes measured in the near future and potential outcomes are stable in expectation. Actual outcomes after treatment are compared with measured outcomes pre-treatment plus predicted treatment effect.

\textbf{Graph probability of negative outcome} --- for different groups, graph the probability of a negative treatment effect.

\textbf{PCA on most important variables} --- conduct a principal component analysis on variables and predicted treatment effects.

\textbf{Q-Score metric} --- the Q-Score measures how well the CATE estimates outperform the ATE on test data.

\textbf{Qini curve} --- Qini curve compares the performance of individual estimates to allocate treatment compared to a random allocation of treatment.

\textbf{Slope test} --- this uses an out-of-sample test of causal forest results to construct a number of linear models and the distribution of parameters across these models can distinguish noise from useful capturing of treatment effect heterogeneity.

\textbf{Rank-weighted Average Treatment Effect} --- ranks the sample in order of estimated treatment effect and then calculates a curve showing the expected treatment effect of just treating the top $p$\% of units.

\end{document}